\documentclass{article}
\usepackage{arxiv}
\usepackage[numbers,sort&compress]{natbib}
\usepackage[utf8]{inputenc} 
\usepackage[T1]{fontenc}    
\usepackage{hyperref}       
\usepackage{url}
\usepackage{textcomp}
\usepackage{ gensymb }
\usepackage[ruled,vlined]{algorithm2e}
\usepackage{algorithmic}
\usepackage{booktabs}       
\usepackage{amsfonts}       
\usepackage{nicefrac}       
\usepackage{microtype}      
\usepackage{lipsum}
\usepackage{float}
\usepackage{graphicx}
\usepackage{multirow}
\usepackage{array}
\usepackage{comment}
\usepackage{longtable}
\usepackage{ragged2e}
\usepackage{subfig}

\title{Image Data collection and implementation of deep learning-based model in detecting Monkeypox disease using modified VGG16}

\author{
  Md Manjurul Ahsan \\
  Industrial and Systems Engineering\\
  University of Oklahoma\\
  Norman, Oklahoma-73071 \\
  \texttt{ahsan@ou.edu} \\
   \And
 Muhammad Ramiz Uddin \\
  Dept. of Chemistry and Biochemistry\\
  University of Oklahoma\\
  Norman, Oklahoma-73019\\
  \texttt{muhammadramizuddin@gmail.com}\\
  \And
 Mithila Farjana \\
  Dept. of Chemistry and Biochemistry\\
  University of Oklahoma\\
  Norman, Oklahoma-73019\\
  \texttt{mithilafarjana@ou.edu}\\
  \And
 Ahmed Nazmus Sakib \\
  Dept. of Aerospace and Mechanical Engineering\\
  University of Oklahoma\\
  Norman, Oklahoma-73019\\
  \texttt{nazmus.sakib@ou.edu}\\
  \And
 Khondhaker Al Momin  \\
  Dept. of Civil Engineering\\
  Daffodil International University\\
  Dhaka, Bangladesh-1341\\
  \texttt{ momin.ce@diu.edu.bd}\\
  \And
 Shahana Akter Luna \\
  Medicine \& Surgery\\
  Dhaka Medical College \& Hospital\\
  Dhaka, Bangladesh- 1000\\
  \texttt{shahanaakterluna123@gmail.com}\\} 

\begin{document}
\maketitle

\begin{abstract}
While the world is still attempting to recover from the damage caused by the broad spread of COVID-19, the monkeypox virus poses a new threat of becoming a global pandemic. Although the Monkeypox virus itself is not deadly and contagious as COVID-19, still every day, new patients case has been reported from many nations. Therefore, it will be no surprise if the world ever faces another global pandemic due to the lack of proper precautious steps. Recently, Machine learning (ML) has demonstrated huge potential in image-based diagnoses such as cancer detection, tumor cell identification, and COVID-19 patient detection. Therefore, a similar application can be adopted to diagnose the Monkeypox related disease as it infected the human skin, which image can be acquired and further used in diagnosing the disease. However, there is no publicly available Monkeypox dataset that can be used for training and experimenting the ML model development. Consequently, there is an immediate need to develop a dataset containing Monkeypox infected patients' images. Considering this opportunity, in this work, we introduce a newly developed "Monkeypox2022" dataset that is publicly available to use and can be obtained from our shared GitHub repository. The dataset is created by collecting images from multiple open-source and online portals that do not impose any restrictions on use, even for commercial purposes, hence giving a safer path to use and disseminate such data when constructing and deploying any type of ML models. Further, we propose and evaluate a modified VGG16 model, which includes two distinct studies: Study One and Two. Our exploratory computational results indicate that our suggested model can identify Monkeypox patients with an accuracy of $97 \pm 1.8\%$ (AUC = 97.2) and $88 \pm 0.8\%$ ( AUC = 0.867)  for Study One and Two, respectively. Additionally, we explain 
our model's prediction and feature extraction utilizing Local Interpretable Model-Agnostic Explanations (LIME) helps to a deeper insights of specific features that characterize the onset of the Monkeypox virus.
\end{abstract}

\keywords{Deep learning \and Disease diagnosis \and Image processing \and Monkeypox virus \and Machine learning \and monkeypox dataset \and Transfer learning}

\section{Introduction}
The advent of the Monkeypox in 2022, reported by many nations, demonstrates another challenge worldwide as the world was affected due to the onset of COVID-19 in 2020. Monkeypox is an infectious disease caused by the Zoonotic Orthopoxvirus, which is closely related to both cowpox and smallpox belongs to the Poxviridae family (a member of the genus Orthopoxvirus)~\cite{mccollum2014human}. It is mostly transmitted by monkeys and rodents; nevertheless, human-to-human spread is also extremely prevalent~\cite{alakunle2020monkeypox}. The virus was first identified in a monkey's body in 1958 in a laboratory in Copenhagen, Denmark~\cite{Monkeypox2022}. The first human case of monkeypox was recorded in the Democratic Republic of the Congo in 1970, during an intensified campaign to eradicate smallpox~\cite{nolen2016extended}. Monkeypox is usually exposed in the central and western part of Africa and affects many individuals who reside near to the tropical rainforests. The virus itself contaminates when a person comes in close contact with another infected person, animal, or material. It is transmitted through direct body contact, animal bites, respiratory droplets, or mucous of the eye, nose, or mouth~\cite{nguyen2021reemergence}. Some of the early-stage symptoms of patients infected with Monkeypox include fever, body aches, and fatigue, wherein the long-term effect has a red bump on the skin~\cite{cdcsymptoms}.

Although Monkeypox is not significantly contagious compared to COVID 19 as reported so far, still, the cases continue to rise. There were only 50 monkeypox cases in 1990 in West and Central Africa~\cite{africacase50}. However, the cases rose to  5000 in 2020. Monkeypox is claimed to occur only in Africa in the past, wherein in 2022, the identification of the individuals infected by the virus is reported by several other non-African countries in the Europe and United States~\cite{multi2022}. As an effect, tremendous anxiety and fear among the people are slowly growing, often reflected through the individual’s opinion on social media.

Currently, there is no appropriate treatment for the Monkeypox virus, according to the guidelines provided by the Centers for Disease Control and Prevention (CDC)~\cite{notreatment2022}. Nevertheless, to cope with the urgent need, the CDC approved two oral drugs, Brincidofovir and Tecovirimat, which have mainly been used to treat the smallpox virus, have now been used to treat the monkeypox virus~\cite{adler2022clinical}. 
However, vaccination is the ultimate solution to the Monkeypox virus. Although vaccines are available for the monkeypox virus approved by the food and drug administration (FDA), they have not yet been used in the United States for human use. In other countries, the vaccines for the smallpox virus are used to treat the monkeypox virus~\cite{vaccineavai2022}. 

The diagnosis procedure of the Monkeypox includes initial observations of the unusual characteristics of skin lesions present and the existing history of exposure. However, the definitive way to diagnose the virus is to test skin lesions using electron microscopy. In addition, the Monkeypox virus can be confirmed using polymerase chain reaction (PCR)~\cite{Monkeypoxagri2022} , which is currently being used extensively in diagnosing the COVID-19 patients~\cite{ahsan2020covid,ahsan2020deep,ahsan2021detecting,ahsan2021detection}.

Machine learning (ML) is an emerging discipline of Artificial Intelligence (AI) with established promises in a number of domains, with facilities ranging from decision-making tools to industrial sectors to medical imaging and disease diagnosis. With the specific qualities of ML, clinicians can obtain secure, precise, and fast imaging solutions, which have acquired widespread acceptance as a useful decision-making tool~\cite{ahsan2022machine}. For instance, Miranda and Felipe (2015) developed computer-aided diagnosis (CAD) systems using fuzzy logic to diagnose breast cancer. Fuzzy logic has the advantage over conventional ML as it can reduce time computational issues while replicating the reasoning and style of a skilled radiologist. If the user assigns factors such as contour, density, and form, the algorithm returns a cancer diagnosis result depending on the preferred technique~\cite{miranda2015computer}. Ardakani et al. (2020) evaluated ten different deep learning models using a tiny dataset containing 108 patients with COVID-19 and 86 non-COVID-19 and reached a 99\% accuracy~\cite{ardakani2020application}. Wang et al. (2020) constructed a modified inception-based model using 453 CT scan images and attained an accuracy of 73.1\%~\cite{wang2020covid}. Sanddep et al. (2022) proposed a low complex convolutional neural network (CNN) to detect skin diseases such as Psoriasis, Melanoma, Lupus, and chickenpox. They show that using exiting VGGNet, it is possible to detect skin disease 71\% accurately using image analysis.
In comparison, their proposed solution demonstrates the best results by achieving an accuracy of around 78\%. Velasco et al. (2019) proposed a smartphone-based skin disease identification utilizing MobileNet and reported around 94.4\% accuracy in detecting patients with chickenpox symptoms~\cite{sandeep2022diagnosis}. Roy et al. (2019) utilized different segmentation approaches to detect skin diseases such as acne, candidiasis, cellulitis, chickenpox, etc~\cite{roy2019skin}.

At the time of writing, not a single research study has been discovered that indicates the potential of ML approaches in the diagnosis of Monkeypox disease utilizing image processing techniques. We identified two primary reasons for the absence of a foundation for the development of an image-based diagnosis of Monkeypox disease:
\begin{enumerate}
    \item There is no publicly accessible dataset that can be utilized to train and test to construct a machine learning (ML) model for diagnosing monkeypox. 
    \item Since the virus is only recently heavily exposed in many nations, it is common for an appropriate ML algorithm to be included with the dataset and a model require more time to present.

\end{enumerate}

Taking these potentials into account, we determined that there is a need to establish a dataset including images of patients with Monkeypox disease, which will enable many researchers and practitioners to immediately begin work on developing and proposing a unique AI-assisted strategy. Our motivation in establishing the Monekypox data set is inspired by Dr. Joseph Cohen, who generated the dataset during the onset of COVID-19 by gathering the dataset from numerous sources, including websites and papers~\cite{cohen2020covid}. Their initial dataset comprised COVID-19 and Non-COVID-19 chest X-ray images of 98 samples, whereas our starting dataset consists of 164 samples, and the total sample size after data augmentation is 1915. Numerous research conducted at the onset of COVID-19 employ limited datasets and highlight the value of transfer learning methodologies by offering numerous deep learning-based screening models~\cite{ahsan2020deep, ardakani2020application,wang2020covid}. Therefore, we anticipate that our dataset will serve the same function and assist researchers and practitioners who are eagerly waiting to get access to the dataset in order to construct a model for diagnosing Monkeypox disease. Our technical contribution is outlined below:
\begin{enumerate}
    \item Develop the first publicly Monkeypox image dataset by collecting images from various sources (i.e., news media, websites) and can be obtained from the following \href{https://github.com/mahsan2/Monkeypox-dataset-2022}{github repository};
    \item  Introduce a low-modified VGG16 model for detection of Monkeypox patients using image data; and
    \item Provide post-image analysis explanation using local interpretable model-agnostic explanations (LIME) to validate our findings.
\end{enumerate}
The remaining paper is structured as follows: Section~\ref{methods} provides a concise explanation of the experiment's methodology, followed by Section~\ref{observation}'s results. Section~\ref{discussion} briefly discusses our study, while Section~\ref{conclusions} summarizes our overall findings with possible future research directions.
\section{Methodology}\label{methods}
This section describes the data collecting and augmentation technique, the development of the proposed modified VGG16 deep learning model, the experimental setup, and the performance assessment matrices used to conduct the experiment.
\subsection{Data collection}
\begin{figure}
    \centering
    \includegraphics[width=\textwidth]{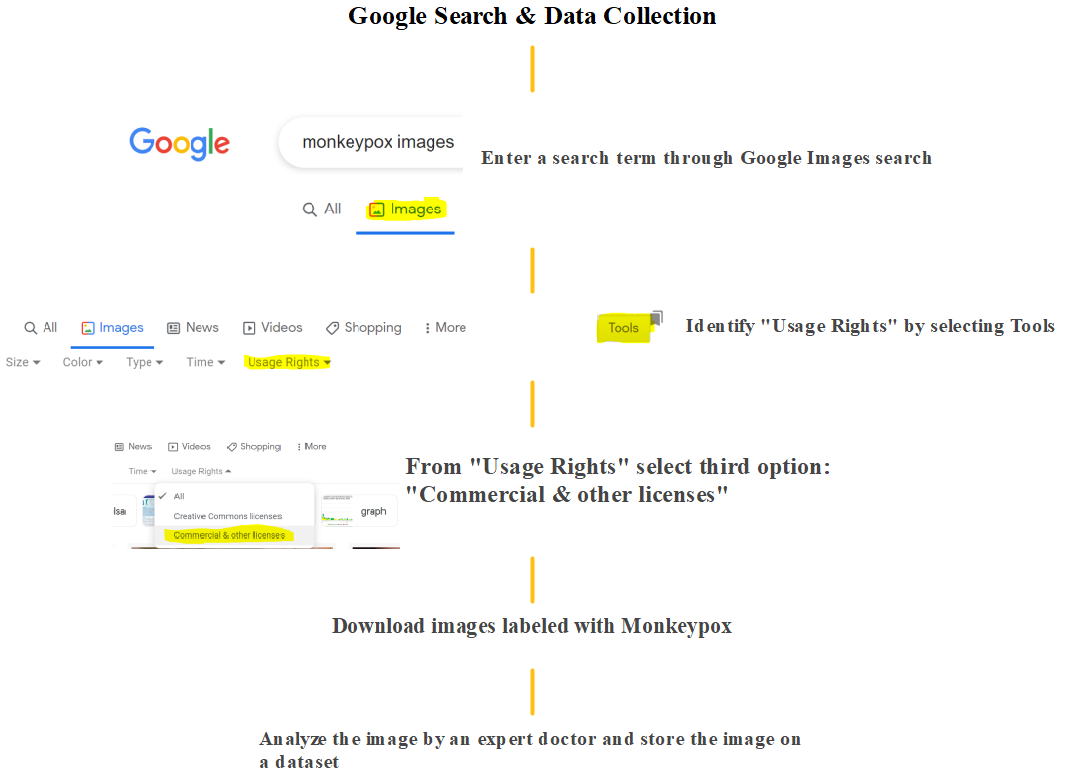}
    \caption{Data collection procedure used in this study.}
    \label{fig:datacolect}
\end{figure}

In the time of the rapidly emerging Monkeypox disease among many nations, it is imperative to diagnosis patients with Monkeypox symptoms. Many experts in the medical domain believes that artificial intelligence (AI) systems could reduce the burden on clinical diagnosis with the outbreaks by processing image data~\cite{ahsan2020covid}. During the onset of COVID-19, we observed that, hospitals in China and Italy, deployed AI-based and image processing based interpreters to improve the hospitals efficiency in handling COVID-19 patients~\cite{ahsan2020deep,ahsan2021detection,narin2021automatic}. However, at the time of writing, unable to find any publicly available Monkeypox dataset hinders taking advantage of deploying an AI-based approach to diagnose and prevent the Monkeypox disease efficiently. As an effect, many researchers and practitioners cannot contribute to detecting Monkeypox disease using advanced AI techniques. Considering these limitations, we collected patients images with Monkeypox images in this work. Our initial dataset contains very limited samples which will not be an issue for the initial experimentation, as supported by many referenced literature that previously considered limited dataset in developing AI-based model during the onset of COVID-19 diseases, refer to~\cite{ardakani2020application,narin2021automatic}. However,  the database will be regularly updated with data contributed by numerous global entities. We followed following procedure to collect the data samples.
\begin{enumerate}
    \item As there is no established shared dataset is available by the authorized and designated hospital, clinic, or viable source, therefore, to establish a preliminary dataset, the Monkeypox image data is collected from various sources such as websites, newspapers, and online portals and publicly shared samples. To do so, the google search engine is used for the initial searching procedure. Figure~\ref{fig:datacolect} displays the procedure used to search the data.
    \item To develop the non-Monkeypox samples, a similar procedure is used in collecting the data sample, which contains search terms such as  “Chickenpox,” “Measles,” and normal images (i.e., photos of both hands, legs, and faces) without any symptoms of the designated disease.
    \item To increase the data sample size, additional Normal images are collected manually from various participants with their consent who do not have any skin disease symptoms. A consent form is used to get approval from all the participants.
\end{enumerate}
Table~\ref{tab:datacollection} summarizes the characteristics of the various datasets developed throughout this study. While transfer learning and conventional machine learning can perform well with a small number of images, deep architecture such as deep learning networks, CNN, RNN and generative adversarial networks require a significant ammount of data samples to construct a model. In consideration of this, we have utilized data augmentation techniques to expand the data set.
\begin{table}[ht]
\caption{The sample size of each dataset that has been collected in this study.}
    \centering
    \begin{tabular}{ll}\toprule
        Dataset&	Total sample\\\midrule
Monkeypox&	43\\
Chickenpox&	47\\
Measles&	17\\
Normal&	54\\
Monkeypox augmented&	587\\
Chickenpox augmented&	329\\
Measles augmented&	286\\
Normal augmented&	552\\\midrule
Total samples&	1915\\\bottomrule
 
    \end{tabular}
    
    \label{tab:datacollection}
\end{table}

Although the dataset contains only 1915 samples, using the traditional ML and transfer learning approach, it is possible to develop a disease diagnosis model as previously demonstrated by many studies during the onset of COVID-19 when the data samples were very limited. For instance, some study uses only 40-100 samples and develop deep learning models to classify COVID-19 patients~\cite{narin2021automatic,ahsan2020covid}. Therefore, we can assuem that our dataset at the very early stage of the Monkeypox is more reliable as the dataset contains around 1915 samples. However, we expect that the dataset will expand over the time as we will collect more data from various open-source (i.e., data available to use without privacy concerns, data from the journals and online).
\subsection{Data augmentation}
Keras image processing library such as ImageDataGenerator is used to augment the dataset. ImageDataGenerator function provides various options such as rotation, width and height shifting, and flipping. A details facility provided by ImageDataGenerator can be found in~\cite{tensorflow2020}. In this work following parameter is used to augment the image data as shown in Table~\ref{tab:aug}. The generator type and facility types are selected randomly as suggested in~\cite{sreene2020}.
\begin{table}[ht]
   \caption{Data augmentation techniques used in this study.}
    \centering
    \begin{tabular}{ll}\toprule
        Generator Type&	Facility\\\midrule
        Width shift&	Up to 2\%\\
Rotation Range&	Randomly 0\degree-45\degree\\
Zoom range&	2\%\\
Height shift&	Up to 2\%\\
Shear range&	2\%\\
Fill mode&	Reflective\\
Horizontal flip&	True\\\bottomrule

    \end{tabular}
    \label{tab:aug}
\end{table}

Algorithm~\ref{dataaug} shows the pseudocode for data augmentation techniques used in this study.
 \begin{algorithm}
\caption{Pseudo-Code of Data Augmentation}\label{dataaug}
\begin{algorithmic}[h]
\STATE Input: read original image samples x using OpenCV.
\STATE  Resize image into $128 \times 128$. 

\STATE  Store resize image as an array inside a list.

\STATE  Call Image data generator function
 \FOR{$n \gets 1$ to 20}
\STATE Batch size = 16
\STATE Save to directory
\STATE Save format as “png”
\ENDFOR
\STATE End of Pseudo-Code.
\end{algorithmic}
\end{algorithm}

\begin{figure}
    \centering
    \includegraphics[width=.5\textwidth]{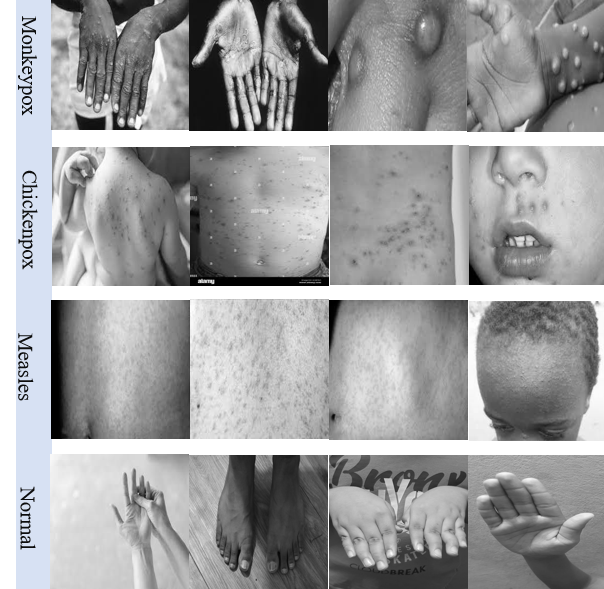}
    \caption{Sample set of images from the developed dataset including Monkeypox, Chickenpox, Measles, and normal images.}
    \label{fig:datasample}
\end{figure}

Figure~\ref{fig:datasample} displays sample images of various datasets developed throughout this study.
\subsection{Transfer learning approaches}
\begin{figure}
    \centering
    \includegraphics[width=\textwidth]{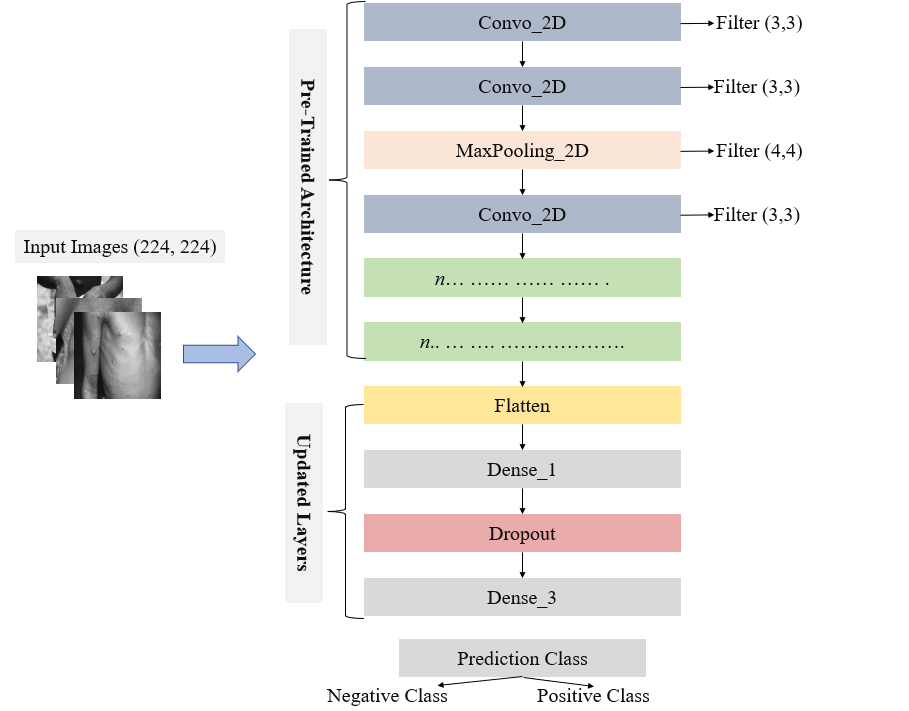}
    \caption{VGG16 models implemented using transfer learning approaches during this experiment.}
    \label{fig:vgg}
\end{figure}

Initially, to evaluate the performance of ML models on developed dataset transfer learning approach is applied for the pilot test. In the preliminary experimentation, a modified VGG16 model is used~\cite{qassim2018compressed}. The core model consists of three essential elements: pre-trained architecture, an updated layer, and a prediction class (partially adapted from~\cite{ahsan2020covid}). The pre-trained architecture is used to identify high-dimensional features and is further added to the updated modified layer. Figure~\ref{fig:vgg} illustrates the proposed updated VGG16 models. The model consists of sixteen CNN layers with varying filter sizes and stride values~\cite{ahmed2020impact}. As shown in Figure, after the initial input layer (consider $224 \times 224$ images only), two convolutional layer (containing a 3x3 filter) is added, followed by a Max Pooling layer, followed by another two convolutional and one Max Pooling layer until it reaches to the modified layer sections. The modified layer Flattened the architecture, followed by the three dense and one dropout layer.

The batch size, number of epochs, and learning rate are initially examined during parameter tuning to maximize the performance of the proposed model. The following experiment parameters are selected at the beginning for Study One (inspired by~\cite{ahsan2021detecting,bergstra2012random}):\\
\begin{center}
Batch size = [5, 10, 15, 20]\\
Learning rate = [0.1, 0.01, 0.001]\\
Number of Epochs = [30, 35, 40, 45]\\   
\end{center}
Using the grid search method following parameters are identified as the most optimal ones:\\
\begin{center}
Batch size = 10\\
Learning rate = 0.001\\
Number of Epochs = 35\\   
\end{center}
For Study Two following parameters are used to develop the optimal model:
\begin{center}
Batch size = [30, 50, 70, 100]\\
Learning rate = [0.1, 0.01, 0.001]\\
Number of Epochs = [50, 70, 100, 150]  
\end{center}
And the best result was achieved with:
\begin{center}
Batch size = 30\\
Learning rate =  0.001\\
Number of Epochs =  100\\   
\end{center}
An adaptive momentum estimation (Adam) is used to optimize the model loss due to its robust performance compared to other exiting optimize algorithms. One of the potential advantages of Adam is that it often demonstrates outstanding performance in binary image classification~\cite{perez2017effectiveness}.
\subsection{LIME as explainable AI}
Local interpretable model-agonistic explanations (LIME) is one of the powerful tools that can help to analyze the model's true prediction and offer the opportunity to understand the Blackbox behind any CNN model's final predictions~\cite{ribeiro2016should}. LIME's impressive performance in describing the complexities of image classification has led to its extensive application in recent years~\cite{cian2020evaluating}. In the case of image classification, LIME uses superpixel. When an image is over-segmented, superpixels are produced. Superpixels stores much data and help to identify essential features of the images during the primary prediction~\cite{ahsan2020covid}. Table~\ref{tab:lime} represents the LIME parameters that have been used in this study to calculate the superpixel values. Note that the parameters are proven to be useful in many image prediction analyses, as referred to in many existing literatures~\cite{ahsan2020covid,pan2020prognostic}. 

\begin{table}[ht]
\caption{Parameter used to identify superpixels.}
    \centering
    \begin{tabular}{ll}\toprule
    Function& Value\\\midrule
    Maximum distance&	150\\
         	Kernel size&	4\\
Ratio&	0.2\\\bottomrule

    \end{tabular}
    \label{tab:lime}
\end{table}
\subsection{Experiment setup}
 The experiment was conducted using a traditional laptop within the specification of Windows 10, 16 GB RAM, and Intel Core I7. The overall experiment was run five times, and the final result is presented by averaging all the five computational outcomes.

Table~\ref{tab:exdata} summarizes the dataset used during this study. The table shows that the dataset contains 43 Monkeypox and 47 Chickenpox images for Study One, whereas the Study Two dataset contains 587 Monkeypox augmented and 1167 other samples, respectively. Note that the "Other" class is a combination of the data samples of Chickenpox, Measles, and Normal images. We allocated 80\% of the sample data to train and the rest of the 20\% to test the model, which is a common practice in ML domains~\cite{mohanty2016using,menzies2006data,stolfo2000cost}.
\begin{table}[ht]
\caption{Assignment of data employed to train and test the proposed modified VGG16 deep learning models.}
    \centering
    \begin{tabular}{lllll}\toprule
         Study&	Label&	Train set&	Test set&	Total\\\midrule
\multirow{2}{*}{Study One}&	Monkeypox&	34&	9&	43\\
&Chickenpox&	38&	9&	47\\\cmidrule{2-5}
&Total&		72&	18&	90\\\midrule
				
\multirow{2}{*}{Study Two}&	Monkeypox augmented&	470&	117&	587\\
	&Others&	933&	234&	1167\\\cmidrule{2-5}
&Total& 1403& 351& 1754\\\bottomrule

    \end{tabular}
    
    \label{tab:exdata}
\end{table}
\subsection{Performance evaluation}
The overall experimental outcome is measured and presented using the most widely used statistical approaches such as accuracy, precision, recall, F1-score, sensitivity, and specificity. For Study One, due to the limited samples, the overall statistical results are represented with a 95\% confidence interval followed by previously reported literature that also used a small dataset~\cite{narin2021automatic,wang2020covid}.  In our dataset, Monkeypox might be classified as true positive ($T_p$) or true negative ($T_n$) if individuals are diagnosed accurately, and it might be classified into false positive ($F_p$) or false negative ($F_n$) if misdiagnosed. The designated statistical metrics are explained in details below.

\textbf{\textit{Accuracy:}} The accuracy is the overall number of successfully identified instances across all cases. Using the following formulas, accuracy can be determined:
\begin{equation}
    Accuracy = \frac{T_{p}+ T_{n}}{T_{p}+T_{n}+F_{p}+F_{n}}
\end{equation}
\textbf{\textit{Precision:}}Precision is assessed as the ratio of accurately predicted positive outcomes out of all expected positive outcomes.
\begin{equation}
    Precision = \frac{T_p}{T_p+F_p}
\end{equation}
\textbf{\textit{Recall:}} Recall refers to the ratio of relevant outcomes that the algorithm accurately identifies.
\begin{equation}
   Recall = \frac{T_p}{T_n+F_p}
\end{equation}
\textbf{\textit{Sensitivity:}}
Sensitivity refers to the only accurate positive metric relative to the total number of occurrences and can be measured as follows:
\begin{equation}
    Sensitivity = \frac{T_p}{T_p+F_N}
\end{equation}
\textbf{\textit{Specificity:}}
It identifies the number of accurately identified and calculated true negatives and can be found using the following formula:
\begin{equation}
    Specificity = \frac{T_N}{T_N+F_P}
\end{equation}
\textbf{\textit{F1- score:}} The F1 score is the harmonic mean of precision and recall. The maximum possible F score is 1, which indicates perfect recall and precision.
\begin{equation}
 F1- score = 2\times\frac{\textrm{Precision}\times\textrm{ Recall}}{\textrm{Precision~+~Recall}}
\end{equation}
\textit{\textbf{Area under curve:}} The area under the curve (AUC) represents the behavior of the models under various conditions. AUC can be calculated suing following formulas:
\begin{equation}
    AUC= \frac{\sum r_i(x_p)- x_p((x_p +1)/2}{x_p+x_n}
\end{equation}
Where  $x_p$ and $x_n$ represent positive and negative samples of data, respectively, and $r_i$ represents the rating of the $i${th} positive sample.
\section{Results}\label{observation}
Table~\ref{tab:resultabl} presents the overall accuracy, precision, recall, F1 score, sensitivity, and specificity scores derived from the preliminary computations performed on the train and test set for both Study One and Two. 
In Table~\ref{tab:resultabl}, bold font denotes the best performance for a particular measurement included in this investigation. 
As the dataset only contained a small number of samples, the findings are presented with confidence intervals (CI) of 95\% in order to provide an accurate overview of the statistical measurements. From the table, it is clear that the performance of Study One is significantly higher compared to the Study Two. For instance, the accuracy of the train set in Study One is higher up to 9\% compared to the accuracy of the train set result of Study Two. The high sensitivity score and low specificity score also demonstrates the models’ lower performance on the test set for both Study. However, overall model performance still can be considered satisfactory with such a limited dataset.

\begin{table}[ht]
\caption{Proposed model's performance on the referenced dataset used in this study, along with confidence interval ($\alpha = 0.05$).}
    \centering\resizebox{\textwidth}{!}{
    \begin{tabular}{llllllll}\toprule
         Study& Dataset&	Accuracy&	Precision&	Recall&	F1-score&	Sensitivity&	Specificity\\\midrule
         \multirow{2}{*}{Study One}& Train set&	\textbf{0.97 $\pm$ 0.018} &	\textbf{0.97 $\pm$ .018}&\textbf{0.97 $\pm$ 0.018}&	\textbf{0.97 $\pm$ 0.018}&	0.973 $\pm$ 0.017&\textbf{0.97 $\pm$ 0.018}\\
         & Test set&	0.83$\pm$ 0.085&	0.88 $\pm$ 0.072&	0.83 $\pm$ 0.085&	0.83 $\pm$ 0.85&	\textbf{1}&	0.66 $\pm$ 0.12\\
         \multirow{2}{*}{Study Two}& Train set&	0.88 $\pm$ 0.008&	0.86 $\pm$ 0.009&	0.87$\pm$ 0.008&	0.86$\pm$ 0.009&	0.83$\pm$ 0.010&	0.89$\pm$ 0.008\\
         & Test set&	0.78 $\pm$ 0.022&	0.75$\pm$ 0.023&	0.75$\pm$ 0.023&	0.75$\pm$ 0.023&	0.650$\pm$ 028&	0.83$\pm$ 0.019\\\bottomrule
         
    \end{tabular}}
    \label{tab:resultabl}
\end{table}

\begin{figure}[ht]
    \centering
    \includegraphics[width=\textwidth]{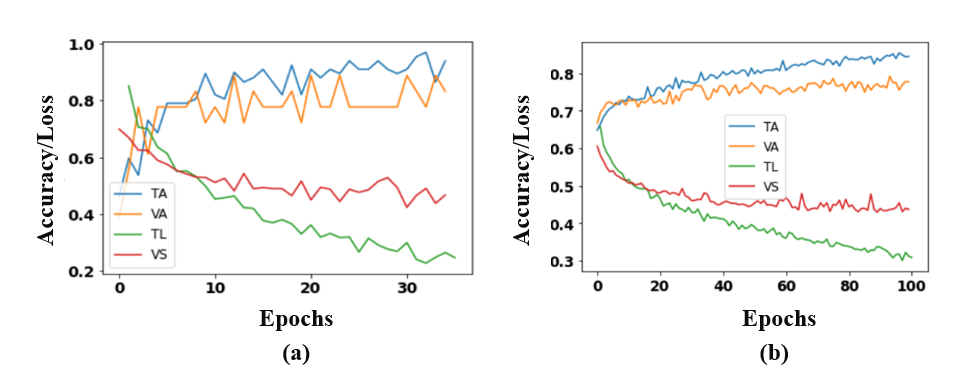}
    \caption{ Modified VGG16 models accuracy and loss during each epoch applied to (a) Study One and (b) Study Two; TA—train accuracy, VA—validation accuracy, TL—train loss, VS—validation loss.}
    \label{fig:al}
\end{figure}

Figure~\ref{fig:al} presents the model performance on the training and testing set per each epoch during each study. In this case, the accuracy reached its highest point before the model started to overfit after 35 epochs for Study One, as shown in Figure~\ref{fig:al}(a). Similarly, the training and validation loss decreases up to 35 epochs, as shown in Figure~\ref{fig:al}(a). In Figure~\ref{fig:al}(b), the model’s performance of Study Two reached its peak point at 100 epochs for both the training and validation dataset.

\begin{figure}[ht]
    \centering
    \includegraphics{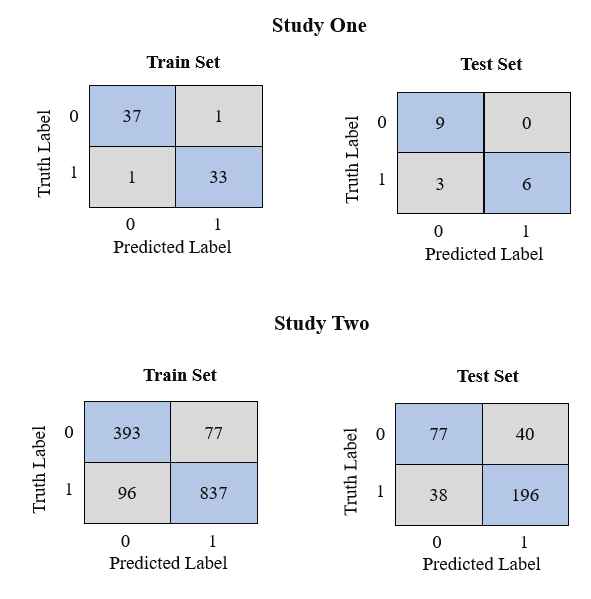}
    \caption{Confusion matrices for the proposed model applied to Study One and Two.}
    \label{fig:cm}
\end{figure}

Figure~\ref{fig:cm} illustrates the proposed model's performance for Study One and Two in terms of the confusion matrix. The figure shows that the lowest misclassification occurs for the train set in Study One (an error rate of 2.7\%). Contrary, maximum misclassification took place for the train set in Study Two (an error rate of 12.33\%). One of the main reasons for achieving the highest number of misclassification during Study Two could be the imbalanced ratio of the dataset, wherein Monkeypox: Others = 1:1.98.

\begin{figure}
    \centering
    \includegraphics[width=.9\textwidth]{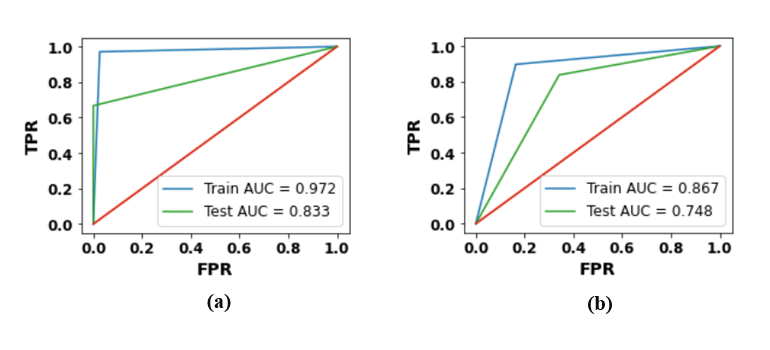}
    \caption{AUC-ROC curves for (a) Study One, and (b) Study Two wherein, TPR—true positive rate, FPR—false positive rate.}
    \label{fig:ar}
\end{figure}

Figure~\ref{fig:ar} represents the AUC of the receiver operator characteristic (ROC) of modified VGG16 models with the true positive rate (TPR) and false positive rate (FPR). The best performance is observed on the train set For Study One by achieving the AUC score of 0.972, as shown in Figure~\ref{fig:ar}(a). Contrary, the worst performance is observed on the test set for Study Two by acquiring the AUC score of 0.748 (refer to Figure~\ref{fig:ar}(b)). Note that AUC score lower than 0.5 indicates the poor performance of the model while a score closer to 1 demonstrates maximum performance. In this regard, the AUC score of 0.748 on the test set also supports the model’s stability during the prediction. 

\begin{figure}
    \centering
    \includegraphics[width=.7\textwidth]{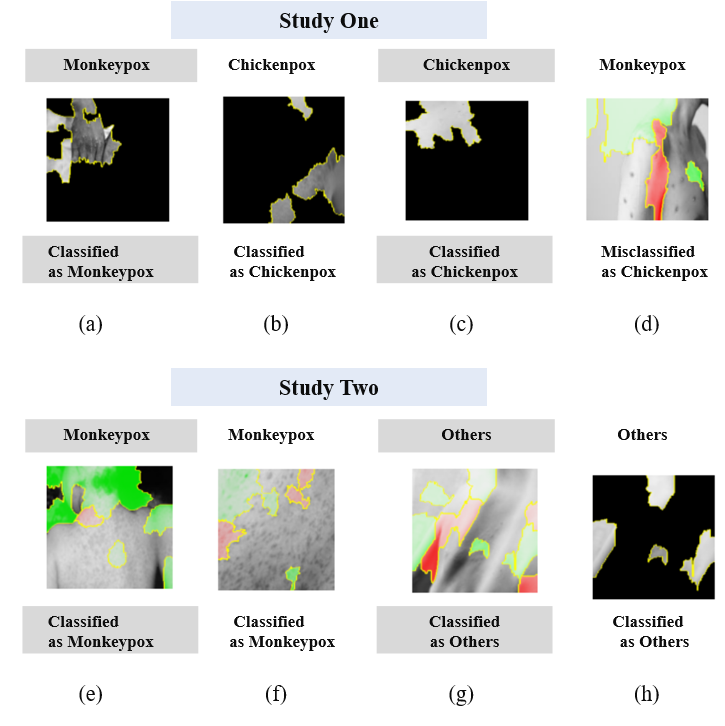}
    \caption{Top four features that enabled the classification between Monkeypox, Chickenpox and Others from the image data.}
    \label{fig:lime}
\end{figure}

Figure~\ref{fig:lime} demonstrates the top four features identified by the proposed models using LIME. From the figure, it can be observed that, for Study One,  in Figure~\ref{fig:lime}(d), the image containing Monkeypox symptoms was misclassified as Chickenpox. Using LIME, from Figure~\ref{fig:lime}(d), we can also evaluate why the proposed model misclassified the images. For instance, here in Figure~\ref{fig:lime}(d), the model mainly focused on the area with no signs or symptoms of Monkeypox. In Study Two, our proposed model accurately classified all four images from the test set, as shown in Figure ~\ref{fig:lime}(e)-(h).
\section{Discussions}\label{discussion}
In this study, we develop a new dataset that can be used to train and develop ML models to classify the Monkeypox disease using image analysis techniques. In addition, a modified VGG16 model is developed and its ability to differentiate between patients with and without Monkeypox disease is evaluated in two separate studies. Our proposed model achieved around the accuracy of $0.83 \pm 0.085$ on a very small dataset, which aligned with many previous studies that demonstrate the superiority of the transfer learning approach on small datasets. Further, we tested our proposed model on an imbalanced dataset and achieved an accuracy of $0.78 \pm 0.022$ on the test set. A recent report presented by World Health Organization (WHO) encouraged any ML model needs to provide proper interpretation before being applied to a clinical trial~\cite{world2021ethics}. Considering this necessity, in this work, we have explained and overlooked our post prediction analysis using one of the popular explainable AI techniques, LIME. Using LIME, we demonstrate that our models are capable of learning from the infected regions and localizing those areas. Our data collection procedure and model performance are analyzed by expert doctors who ensure our model's satisfactory performance. Since there is no Monkeypox image dataset there for, we did not find any single study that can be used to compare our model's performance. However, it can be assumed that our new dataset will give the immense opportunity to researchers and practitioners to practice and develop image-based analysis tools for Monkeypox disease diagnosis.

Finally, healthcare professionals can easily adapt our model as it is cost and time effective and does not require extensive PCR or microscopy testing. As an effect, our proposed model provides an opportunity to test in real-time screening of the patients with Monekypox symptoms. Apart from many advantages, the following limitations of our study provide urgent opportunity for additional research:
\begin{enumerate}
    \item Due to the time constraints, the developed dataset contains limited samples.
    \item The data are primarily collected through various open sources instead of any specific hospital or clinical facility.
\end{enumerate}
\section{Conclusion}\label{conclusions}
The study aims to address the ongoing data scarcity related to Monkeypox virus-infected patient images. The dataset is developed by collecting the images from open source and is publicly available to use without any privacy restrictions, ultimately allowing individuals to share and use that data for experiments and even for commercial purposes. Additionally, We conducted two studies considering small and moderate datasets wherein a modified VGG16 model is implemented. Our findings suggest that using transfer learning approaches, the proposed modified VGG16 can distinguish patients with Monekypox symptoms from others in both Study One and Two with accuracy ranging from 78\% to 97\%. Finally, we have used LIME to present the proper explanation of the reason behind our model's prediction, which is one of the current demands in deploying ML models for clinical trials. Our model's predictions were cross-checked by doctors to emphasize that the results could be validated. We intend to emphasize the possibilities of artificial intelligence-based approaches, which might play an essential role in diagnosing and preventing the contamination of the onset of the Monkeypox virus. We hope our publicly available dataset will play an important role and provide the opportunity to the ML researcher who cannot develop an AI-based model and conduct the experiment due to data scarcity. As our proposed model is supported by many previously published literature that uses the transfer learning approach in developing an AI-based diagnosis model, it will also encourage future research and practitioners to take advantage of the transfer learning approach and apply it in clinical diagnosis. Some of the constraints connected with our work can be overcome by updating the dataset by continuously collecting new Monkeypox infected patients images, evaluating the proposed VGG16 model's performance on highly imbalanced data, comparing the performance of our model with other researchers' findings (once available), and deploy our proposed model in developing mobile-based diagnosis tool.
\bibliographystyle{unsrt}  
\bibliography{main}

\end{document}